\documentclass[onecolumn,floatfix,showpacs,aps,nofootinbib]{revtex4}

\usepackage[dvips]{epsfig}
\usepackage[english]{babel}

\usepackage{bbm}
\usepackage{array}
\usepackage{amsmath}
\usepackage{multirow}
\usepackage{amsfonts}
\usepackage{amssymb}
\usepackage{xcolor}
\usepackage{hyperref}
\hypersetup{colorlinks=true,linkbordercolor=blue,linkcolor=blue, citecolor=blue}
\usepackage{subeqnarray}
\usepackage{indentfirst} %primeiro paragrafo com espaço

\makeatletter
\renewcommand*\l@section{\@dottedtocline{1}{1.5em}{2.3em}}
\makeatother
\begin{document}

\title{Effective lagrangian for a mass dimension one fermionic field in curved spacetime}

\author{R. J. Bueno Rogerio$^{1}$} \email{rodolforogerio@feg.unesp.br}
\author{J. M. Hoff da Silva$^{1}$} \email{hoff@feg.unesp.br}
\author{M. Dias$^{2}$} \email{marco.dias@unifesp.br}
\author{S. H. Pereira$^{1}$} \email{shpereira@feg.unesp.br}
\affiliation{ \\$^1$Universidade Estadual Paulista (UNESP)\\Faculdade de Engenharia, Guaratinguet\'a, Departamento de F\'isica e Qu\'imica\\
12516-410, Guaratinguet\'a, SP, Brazil
\\ $^2$Departamento de Ci\^encias Exatas e da Terra,
Universidade Federal de S\~ao Paulo (UNIFESP),\\
09972-270, Diadema, SP, Brazil.}

%\date{\today}

\begin{abstract}
In this work we use momentum-space techniques to evaluate the propagator $G(x,x^{\prime})$ for a spin $1/2$ mass dimension one spinor field on a curved Friedmann-Robertson-Walker spacetime. As a consequence, we built the one-loop correction to the effective lagrangian in the coincidence limit. Going further we compute the effective lagrangian in the finite temperature regime. We arrive at interesting cosmological consequences, as time-dependent cosmological `constant', fully explaining the functional form of previous cosmological models. 
\end{abstract}

\maketitle

\section{Introduction}

The purpose of constructing a quantum fermionic field whose main output is to be neutral with respect to gauge interactions may indeed be faced as a welcome branch of investigation. The first approach in this way resulted in a bottom-up formulation leading to a spin $1/2$ field endowed with mass dimension one \cite{prd72,jcap}. Since its first appearance in the literature, these spinor fields has been explored in many areas, as accelerator physics \cite{fen2,fen3,fen4}, cosmology \cite{co1,co2,co3,co4,co5,co6,co7,co8,co9,co10,co11, hoffsaulo} and mathematically inclined areas as well \cite{rcd,spinorrepresentation,elkop}.

In the early days of mass dimension one spinors, the theory was presented in such a way that a breaking Lorentz term taken part in the spin sums. As a net result the associated quantum field was non-local and a there was preferred axis of symmetry. After all, the theory was shown to be invariant under $SIM(2)$ and $HOM(2)$ transformations \cite{horv}, being then a typical theory carrying the very special relativity symmetries \cite{cohen}. Quite recently, important advances the spinor dual theory has opened the possibility of circumvent the Weinberg no-go theorem, proposing a spinor field of spin $1/2$ endowed with mass dimension one, local, neutral with respect to gauge interactions, and whose theory respect Lorentz symmetries \cite{1305,Ahluwa2,1602}. We should bring back to the scene the canonical Wigner work on the irreducible representations of the Poincar\'e group \cite{Wigner1}. By Poincar\'e group, as usual, it is understood the semi-simple extension of the orthochronous proper Lorentz group encompassing translations. In investigating the irrep's for this case, no particle as a fermion with canonical mass dimension one (i. e. fermions with bosonic traces, in a manner of speaking) is found.  

The situation is different, however, when discrete symmetries are taken into account, i. e., when not only the orthochronous proper group is considered. This point was also analyzed by Wigner, in a less known paper \cite{Wigner2}. Interestingly enough, Wigner found fermionic irrep's whose behavior under\footnote{Being $C$, $P$ and $T$ the usual charge conjugation, parity and time reversal operators, respectively.} $C, P$ and $T$ is exactly what is expected for bosonic (quantum of) fields. For concreteness, while conventional wisdom stay that fermions belonging to the standard model (quarks and leptons) obey $T^2=-1$ ($(CPT)^2=-1$) and bosons $T^2=+1$ ($(CPT)^2=+1$), Wigner shown that, in the very realm of full Poincar\'e symmetries, it is possible to have $T^2=+1$ for fermions (leading also to $(CPT)^2=+1$) for fermions. It turns out that the field taken into account in this work performs a realization of the (indeed odd) aforementioned fermionic representation, from where we can adduce its ``bosonical character''.

In this work we compute the one-loop effective lagrangian for such a spinor field and study its cosmological consequences. The idea is to take advantage of the background-field method developed by DeWitt and Schwinger \cite{dewitt, schwinger1}, applying it to a well behaved curved spacetime by means of the coincidence limit in which momentum space techniques may be used. By studying the effective lagrangian, a time dependent cosmological constant can be identified. A cosmological setup as such can be traced back as a quantum field theory requirement \cite{murat, berto, ozer1, gasperini1, gasperini2, freese}. Most relevant to the present discussion is that a dynamical cosmological constant may provide an adequate framework to seed light on the cosmic expansion, see for instance  \cite{sandro} (and references therein). Usually, the necessary specific behavior of the cosmological constant is believed to be a consequence of a given quantum effect of the universe primordial stages \cite{novello1, novello2, peebles1}. The mechanism presented in this paper is appropriate to describe the cosmological constant behavior since inflationary stages until its current value. The finite temperature quantum corrections were also considered at the end, including a possible cosmological application for recent times, {when a smooth evolution of the background can be supposed}. 

The paper is organized as follows: the next section is reserved for an overview on the formalism endowed to built the one-loop correction effective lagrangian. Section III is devoted to the one-loop corrections. In Section IV we explore some interesting cosmological consequences in two relevant limits. Section V explores the finite temperature effects and its cosmological consequences. In the last section we conclude.  

\section{A short review on the  effective lagrangian formalism}\label{genesis}
The method we shall apply to find the mass dimension one fermionic field effective lagrangian was discussed in many papers over the years \cite{brown,parker,sty,hu,cheng}. In this section we shall illustrate it by means of a scalar field  in a curved spacetime. The lagrangian reads 

\begin{eqnarray}\label{l1}
\mathcal{L} = -\frac{1}{2}\phi(\Box+m^2+\xi R)\phi,
\end{eqnarray}
being $\Box$ the Laplace-Beltrami operator, $R$ the scalar of curvature and $\xi=\frac{1}{4}\frac{(d-2)}{(d-1)}$ the conformal coefficient. In order to evaluate the effective lagrangian, the background-field method should be implemented \cite{dewitt}. Therefore, we split the quantum field excitation from the background (classic) field as   
\begin{eqnarray}
\phi \rightarrow \phi + h,
\end{eqnarray}
where $h$ stands for the quantum field fluctuation and $\phi$ is the classical background \cite{brown}. It is straightforward to see that Eq.\eqref{l1} can be written as
\begin{eqnarray}\label{lh}
\mathcal{L} = -\frac{1}{2}h(\Box+\alpha^2)h,
\end{eqnarray}
where $\alpha^2\equiv m^2+\xi R$ is the so-called effective mass. Consequently, the lagrangian induced by one-loop effects, say $\mathcal{L}^{(1)}$, is given by the functional integral over the quantum fields
\begin{eqnarray}\label{induced}
\exp\bigg(\frac{i}{\hbar}\int dx\mathcal{L}^{(1)}\bigg)= \mathcal{N}\int dh\exp\bigg(\frac{i}{\hbar}\int dx \mathcal{L}\bigg),
\end{eqnarray}
where $ \mathcal{N}$ is just a normalization factor.

Differentiating both sides of Eq.\eqref{induced} with respect to $\alpha^2$, we have
\begin{eqnarray}\label{funcional}
\frac{\partial\mathcal{L}^{(1)}}{\partial\alpha^2}&=&-\frac{1}{2}\frac{\int h(x)h(x^{\prime})exp[(i/\hbar)\int dx \mathcal{L}])dh}{\int exp[(i/\hbar)\int dx \mathcal{L}]dh}\\
&=& \lim_{x\rightarrow x^{\prime}}-\frac{1}{2} \langle h(x)h(x^{\prime})\rangle,
\end{eqnarray} and therefore
\begin{eqnarray}
\frac{\partial\mathcal{L}^{(1)}}{\partial\alpha^2} = \lim_{x\rightarrow x^{\prime}} -\frac{\hbar}{2i}G(x-x^{\prime}),
\end{eqnarray}
where $G(x-x^{\prime})$ is the Green's function that satisfies the equation
\begin{eqnarray}
(\Box+\alpha^2)G(x-x^{\prime}) = (g)^{-1/2}\delta(x-x^{\prime}).
\end{eqnarray} Hereupon, the one-loop effective lagrangian is obtained after integrating the propagator in  $\alpha^2$, in the coincidence limit $x\rightarrow x^{\prime}$. Accordingly, this is the recipe that will form the foundation to compute one-loop correction for mass-dimension-one fermions, to be developed and presented here. 

\section{One-loop corrections to mass dimension one fermionic fields}\label{sec2}
Consider the mass-dimension-one field lagrangian in a curved space-time scenario
\begin{eqnarray}\label{1}
\mathcal{L}_{0}= \frac{1}{2}\sqrt{-g}\big[g^{\mu\nu}(\nabla_{\mu}\stackrel{\neg}{\lambda}\nabla_{\nu}{\lambda})-m^2\stackrel{\neg}{\lambda}\lambda -\xi R\stackrel{\neg}{\lambda}\lambda \big],
\end{eqnarray}
where $\stackrel{\neg}{\lambda}$ and $\lambda$ stands for the adjoint and usual field, respectively \cite{1305}. The metric in a spatially flat, homogeneous and isotropic Friedmann-Lama\^{i}tre-Robertson-Walker (FLRW) expanding universe is given by
\begin{eqnarray}\label{metric}
ds^2 = dt^2 - a^2(t)(dx^2+dy^2+dz^2),
\end{eqnarray}
hereupon
\begin{eqnarray}
g_{\mu\nu} = diag(1,-a^2(t),-a^2(t),-a^2(t)),
\end{eqnarray}
and
\begin{eqnarray}
g^{\mu\nu} = diag(1,-1/a^2(t),-1/a^2(t),-1/a^2(t)).
\end{eqnarray}

The covariant derivatives, $\nabla_{\mu}$, are defined as  $\nabla_{\mu}\stackrel{\neg}{\lambda}= \partial_{\mu}\stackrel{\neg}{\lambda} + \stackrel{\neg}{\lambda}\Gamma_{\mu}$ and
$\nabla_{\mu}\lambda = \partial_{\mu}\lambda - \Gamma_{\mu}\lambda$, where $\Gamma_{\mu}$ denotes the spin connection given by $\Gamma_0=0$ and $\Gamma_j=-\frac{\dot{a}(t)}{2}\gamma_0\gamma_{j}$, where $\gamma_j$ stands for the Dirac matrices
\begin{eqnarray}
\gamma^0(t) = \gamma_{0}, \qquad\mbox{and}\qquad \gamma^{j}(t) = -\frac{1}{a(t)}\gamma_{j},
\end{eqnarray}
where $\gamma_{\mu}$ stands for the Dirac matrices in Minkowsky spacetime in the Weyl representation. 

Towards to execute the background-field method presented in Refs. \cite{brown, parker}, firstly it is necessary to split the field in its classical background $\lambda$ and the quantum fluctuation $\psi$, $\lambda\rightarrow\lambda+\psi$. The one-loop effects, whose net effect here is encoded in $\mathcal{L}^{(1)}$, will be governed by the functional integral over the fields as in Eq.\eqref{induced}. Thus, the one-loop contribution to the effective lagrangian is now related to the Green's function by
\begin{eqnarray}\label{loop}
\frac{\partial\mathcal{L}^{(1)}}{\partial\alpha^2} &=& -\frac{1}{2} \langle \bar{\psi}^{\beta}(x)\psi_{\beta}(x^{\prime})\rangle
\\
&=&\lim_{x\rightarrow x^{\prime}}-\frac{\hbar}{2i}Tr G(x-x^{\prime}),
\end{eqnarray} 
with the subtlety of taking the trace over the spinor indexes \cite{brown}.

The appropriate lagrangian for mass dimension one fermionic fields in the curved space reads
\begin{eqnarray}\label{lagranm}
\mathcal{L}= \frac{1}{2}\sqrt{-g}\big[g^{\mu\nu}(\nabla_{\mu}\stackrel{\neg}{\lambda}\nabla_{\nu}{\lambda})-m^2\stackrel{\neg}{\lambda}\lambda -\xi R\stackrel{\neg}{\lambda}\lambda\big],
\end{eqnarray} where we are ignoring self-coupling terms. The corresponding equation of motion for the quantum fluctuation field can be written as
\begin{eqnarray}
\bigg(\Box+3 H(t)\partial^0+\frac{2}{a^2(t)}\Gamma_{j}\partial^{j} + \alpha^2\bigg)\psi=0,
\end{eqnarray}
where the effective mass is given by $\alpha^2\equiv m^2 +\xi R-\frac{3}{4}H^2(t)$, being $H(t)=\dot{a}/a$ the Hubble parameter. Notice the presence of a first derivative term, coming from the spin sums. Even with a typical scalar field lagrangian, the spinor character of the field at hand shows up its idiosyncrasies. In order to proceed to the Green function, it is necessary to perform a Wick rotation, $t\rightarrow -it$, after what we have 
{
\begin{eqnarray}
\bigg(\partial^2_E+\partial^2_j-3iH(it)\partial^0+\frac{2}{a^2(it)}\Gamma_{j}\partial^{j} + \alpha_{E}^2\bigg)G_{E}(-it,\vec{x}, -it^{\prime}, \vec{x}^{\prime})= (-g)^{-1/2}\delta(-it,\vec{x}, -it^{\prime}, \vec{x}^{\prime}),
\end{eqnarray} 
%$y\equiv x-x^{\prime}$, 
where $\alpha^2_E$ is the Euclidean effective mass. 
%and the Euclidean Green's function is defined as $G_E(y)=G_E(-it,x;-it^{\prime},x^{\prime})$.
 As stated in \cite{hu}, this momentum-space representation is well-defined only in a local neighborhood of $x-x^{\prime}=0$ (or better saying $x\rightarrow x^{\prime}$). However, for the study of ultraviolet divergences in two-point (like Feynman propagator) or bivector quantities (as the energy-momentum tensor) when the coincidence limits are taken, or for the study of systems involving low-order quasilocal variations of the background field (as we present here), results based on the use of the momentum-space representation technique are be valid. It would not be sufficient for the consideration of processes involving rapid changes of the background field, as, for example, pair productions and topological effects due to phase transition. In these situations, the method based in the Heat-Kernel \cite{nepo}, which is a similar method providing the same net result \cite{hu}, would be more profitable since it allows for perturbative treatments of more complicated situations, e.g., when higher derivative of the background field is included.}

{
Now, we shall take advantage of the coincidence limit (the quasi-local situation) to use the momentum space quantization toolkit writing 
\begin{eqnarray}
G(p) = \int dx e^{ip_{\mu}(x-x^{\prime})^{\mu}}G_{E}(-it,\vec{x}, -it^{\prime}, \vec{x}^{\prime}),
\end{eqnarray}
leading, then, to 
\begin{eqnarray}\label{15}
G_E(x\rightarrow x^{\prime}) = \int \frac{d^4p}{(2\pi)^4} \int_0^\infty ds e^{-\big(p^2-3iH(it)p^0+\frac{2i}{a^2(it)}\Gamma_j p^j+\alpha^2_E \big)s}.
\end{eqnarray}
}

Decomposing the momentum usually as  $p^{\mu} = (p^0, \vert p\vert\sin\theta\cos\phi, \vert p\vert\sin\phi\sin\theta, \vert p\vert\cos\theta)$, we have\footnote{As a remark we emphasize that had we working with the first Elko formulation, whose relativist symmetries are governed by $SIM(2)$ or $HOM(2)$ Lorentz subgroups, then the equivalent momentum space Green function would be
\begin{eqnarray}
G_E(x-x^{\prime}) = \frac{1}{(2\pi)^4}\int_0^\infty ds\int d^4p e^{-(p^2-3iH(it)p^0+\frac{2i}{a^2(it)} \Gamma_j p^j+\alpha_E^2 + M_{p})s},
\end{eqnarray} where $M_{p}$ is the momentum space version of a $\phi$-dependent matrix giving rise to a preferred direction. This matrix can be expressed in terms of fractional derivatives \cite{cheng} and its explicit form makes, up to our knowledge, the above integration unviable.} 
\begin{eqnarray}\label{16}
G_E(x\rightarrow x^{\prime}) = \frac{-i}{(2\pi)^4}\int_0^\infty dse^{-\alpha^2_E s} \int_0^\infty dp^0 e^{({-p^0}^{2}+3iH(it)p^0)s}\int_0^{2\pi}d\phi \int_0^\infty dp\vert p\vert^2\int_0^{2\pi}d\theta\sin\theta e^{\big(-\vert p\vert^2+\frac{H(it)}{a(it)}\gamma_0\gamma_j p^j\big)s}.
\end{eqnarray}
Even being Gaussian integrals, the matrices present in the exponential makes the integration a bit laborious. The net result can be written in terms of incomplete gamma functions \cite{abramo, ederlyi}
\begin{eqnarray}\label{17}
G_E(x\rightarrow x^{\prime}) = \frac{-i}{32\pi^2}\int_0^\infty\frac{ds}{s^{2}}e^{-\alpha_E^2s}e^{-\frac{9}{4}H^2(it)s}\bigg[1+\frac{1}{\sqrt{\pi}}\gamma\bigg(\frac{1}{2};-\frac{9}{4}H^2(it)s\bigg)\bigg]\mathbbm{1},
\end{eqnarray} 
where $\mathbbm{1}$ stands for the $4\times 4$ identity matrix.The Euclidean effective one-loop lagrangian is obtained by inserting Eq.\eqref{17} into Eq.\eqref{loop}, after to
switch back from the Euclideanized form, i. e., writing $\mathcal{L}_{E}\rightarrow -\mathcal{L}$ and $H^2(it)\rightarrow -H^2(t)$,  we have
\begin{eqnarray}
\mathcal{L}^{(1)} = \frac{\hbar}{16\pi^2}\int_0^\infty \frac{ds}{s^{3}}e^{-\alpha^2 s}e^{\frac{9}{4}H(t)^2s}\bigg[1+\frac{1}{\sqrt{\pi}}\gamma\bigg(\frac{1}{2};\frac{9}{4}H^2(t)s\bigg)\bigg].
\end{eqnarray}

In the sequel we expand the exponential as $e^{-Us} = \sum_{l=0}^\infty d_{l}s^{l}$, where $d_l\equiv {(-1)}^{l}\frac{(\xi R-3H^2(t))^{l}}{l!}$  and write the incomplete gamma function as a power series \cite{ederlyi}
\begin{eqnarray}
\gamma(z,x) = \sum_{n=0}^{\infty}\frac{(-1)^{n}}{n!}\frac{x^{z+n}}{(z+n)}.
\end{eqnarray} The one-loop lagrangian can thus be written as 
\begin{equation}\label{22}
\mathcal{L}^{(1)} = \frac{\hbar}{16\pi^2}\sum_{l=0}^\infty d_{l}\int_0^\infty dse^{-m^2s}s^{l-3}+
\frac{1}{\sqrt{\pi}}\frac{\hbar}{16\pi^2}\sum_{l=0}^\infty \sum_{n=0}^{\infty}d_{l}\frac{(-1)^{n}}{n!}\frac{\bigg(\frac{9}{4}H^2(t)\bigg)}{(1/2+n)}^{^{1/2+n}}\int_0^\infty dse^{-m^2s}s^{l+n-5/2}.
\end{equation} Taking advantage of the complete gamma function we can express the effective lagrangian as follows
\begin{eqnarray}\label{23}
\mathcal{L}^{(1)} = \frac{\hbar}{16\pi^2}\bigg[\sum_{l=0}^\infty d_{l} {(m^2)}^{2-l}\Gamma(l-2)+\frac{1}{\sqrt{\pi}}\sum_{l=0}^\infty \sum_{n=0}^\infty d_{l}\frac{(-1)^{n}}{n!}\frac{\bigg(\frac{9}{4}H^2(t)\bigg)}{(1/2+n)}^{^{1/2+n}}{(m^2)}^{3/2-l-n}\Gamma\bigg(l+n-3/2\bigg)\bigg].
\end{eqnarray} Aiming to study the cosmological consequences of such quantum corrections, we propose to investigate its impacts on an effective cosmological constant and scalar of curvature \cite{novello1, novello2}. Considering the general gravitational action 
\begin{eqnarray}\label{grav}
\mathcal{S}_{grav}=\int d^4x\mathcal{L}_{grav}= -\frac{1}{16\pi G}\int d^4x \sqrt{-g} (R+2\Lambda),
\end{eqnarray}       
where $\Lambda$ is the cosmological constant and $G$ is the gravitational constant and expanding Eq.\eqref{23} in terms of $l$ and $n$, neglecting terms of orders higher than $H^2(t)$, the total effective lagrangian given by the sum of Eqs.\eqref{1}, \eqref{23} and \eqref{grav}, reads 
\begin{eqnarray}
\mathcal{L}_{eff} &=&\left. \mathcal{L}_{0}+\bigg(\frac{m^4\hbar(3-2\gamma)}{64\pi^{2}}-\frac{3H^2(t)m^2(1-\gamma)\hbar}{16\pi^{2}}+\frac{2H(t)m^3\hbar}{16\pi^{2}}-\frac{2\Lambda}{16\pi G}\bigg)\sqrt{-g}\right. \nonumber\\&+&\left.\bigg(\frac{m^2\xi(1-\gamma)\hbar}{16\pi^{2}}+\frac{1}{16\pi G}\bigg)\sqrt{-g}R + \mathcal{O}(R^2)\right.,
\end{eqnarray}
where $\gamma$ is the Euler-Mascheroni constant. Notice the appearance of the minimal coupling dependence in the $R$ correction.  

\section{Cosmological implications}\label{cosmoscenario}
In order to study some consequences of the above action into a cosmological context, we derive the FRW equations by means of its Lagrangian formulation, which basically consists in introducing a lapse function $N(t)$ into the metric (\ref{metric}) as $ds^2 = N^2(t)dt^2 - a^2(t)(dx^2+dy^2+dz^2)$. The Euler-Lagrange equations obtained by variations of $\mathcal{L}$ with respect to $N(t)$ and $a(t)$ will furnish the two Friedmann equations, and at the end we make $N(t)=1$. We assume that the spinorial field  corresponding to the matter content in $\mathcal{L}_0$ can be split as $\lambda(x)=\phi(t)\chi (\vec{x})$, with $\chi$ satisfying the normalization condition $\stackrel{\neg}{\chi}\chi=1$, a convenient fact justified in Ref. \cite{hoffsaulo}. Moreover, we shall investigate the case in which the spinor field is homogeneously filling all the universe, so that $\nabla_i\chi(\vec{x})=0$, {and also the background evolution is smooth and adiabatic. Such condition is naturally satisfied at late time evolution of the universe}. With these assumptions, the complete Lagrangian in the presence of the $N(t)$ function is\footnote{Tracing back the lapse function presence consequences, it is fairly simple to see that it amounts to be  $\sqrt{-g}=N(t)a^3(t)$, $\Gamma_{\mu}\Gamma^{\mu}=-\frac{3}{4}\frac{\dot{a}^2(t)}{N^2(t)a^2(t)}\mathbbm{1}$ and $\Gamma_{j}=-\frac{\dot{a}(t)}{2N(t)}\gamma_0\gamma_j$. We also have $R=-\frac{6}{N^2(t)a(t)}\bigg(\ddot{a}(t)+{\dot{a}^2(t)\over a^2(t)} - {\dot{a}(t)\dot{N}(t)\over N(t)}\bigg)$.}
\begin{eqnarray}
\mathcal{L}_{eff} &=& -{1\over N}\bigg({3a\dot{a}^2\over 8\pi G}  - {1\over 2}a^3\dot{\phi}^2-{3\over 8}a\dot{a}^2\phi^2 - 3\xi[a\dot{a}^2\phi^2+2a^2\dot{a}\phi\dot{\phi}] +(1-2\xi)\frac{3m^2(1-\gamma)\hbar}{16\pi^2}a\dot{a}^2\bigg)\nonumber\\
&&-Na^3\bigg({1\over 2}m^2\phi^2+{\Lambda\over 8\pi G} - \frac{m^4(3-2\gamma)\hbar}{64\pi^2}\bigg)+\frac{2m^3\hbar}{16\pi^2}\dot{a}a^2\,.
\end{eqnarray} The corresponding Friedmann equations are:
\begin{equation}
H^2=\frac{8\pi G}{3}\bigg[{1\over 2}\dot{\phi}^2 + {1\over 2}m^2\phi^2 + {3\over 8}H^2\phi^2 +{\Lambda\over 8\pi G}-3\xi[H^2\phi^2+2H\phi\dot{\phi}]-\frac{m^4(3-2\gamma)\hbar}{64\pi^2}-(1-2\xi)H^2\frac{3m^2(1-\gamma)\hbar}{16\pi^2} \bigg]\,,\label{H2h}
\end{equation}
\begin{eqnarray}
-2\dot{H}-3H^2&=&8\pi G\bigg[{1\over 2}\dot{\phi}^2 - {1\over 2}m^2\phi^2 - {3\over 8}H^2\phi^2 -{1\over 4}\dot{H}\phi^2-{1\over 2}H\dot{\phi}\phi -{\Lambda\over 8\pi G} +\xi[3H^2\phi^2+4H\phi\dot{\phi}+ 2\dot{H}\phi^2+2\dot{\phi}^2+2\phi\ddot{\phi}]\nonumber\\
&&+(1-2\xi)[H^2+{2\over 3}\dot{H}]\frac{3m^2(1-\gamma)\hbar}{16\pi^2} \bigg]\,.\label{Hph}
\end{eqnarray}

Written in this form we recognize the energy density of the field and its quantum corrections on the right-hand side of (\ref{H2h}) and the pressure and its quantum corrections on the right-hand side of (\ref{Hph}). In the limit $\hbar \to 0$ and $\xi = 0$ we recover the torsion free equations obtained in \cite{sajf}. The last term on the right of (\ref{H2h}) corresponds to the quantum correction to $H^2$ while the term proportional to $m^4\hbar$ is the correction to the cosmological constant term. Within the plethora of research possibilities we shall depict two interesting limits in what follows. 

\subsection{Limit $H\approx m_{pl} \gg m \gg \phi$.}

In order to look for possible consequences of the above equations into early universe, where quantum effects may be relevant, we analyse the first Friedmann equation (\ref{H2h}) in the limit $H\approx m_{pl} \gg m \gg \phi$, which corresponds to an universe of about $t=H^{-1}\sim 10^{-43}$s, before inflation occur. In this limit we have
\begin{equation}
H^2=\frac{8\pi}{3}\bigg[{\dot{\phi}^2\over 2m_{pl}^2}  +{\Lambda\over 8\pi}-(1-2\xi){H^2\over m_{pl}^2}\frac{3m^2(1-\gamma)\hbar}{16\pi^2} \bigg]\,,\label{H2ha}
\end{equation}
where we have introduced the Planck mass $m_{pl}=1/\sqrt{ G}$. If the kinetic term is negligible we see that a positive contribution to the cosmological constant $\Lambda$ can be obtained if $\xi>1/2$, indicating that even in the absence of a cosmological constant term, i. e. $\Lambda=0$, a positive contribution proportional to $H^2$ in the last term survives, which can be interpreted as a cosmological term induced just by quantum effects and could be responsible for the inflationary phase at early universe stages. As the universe expands and $H$ diminish the other terms starts to dominate, as the terms proportional to $\phi$.

\subsection{Limit $\dot{\phi}\ll H\phi$, $\ddot{\phi}\ll H\dot{\phi}$ and $\phi \approx constant$.}

As an application of this spinor field as a candidate to dark energy in late time evolution of the universe, we suppose it as a nearly constant field ($\phi \approx constant$) satisfying a slowly varying condition, $\dot{\phi}\ll H\phi$ and $\ddot{\phi}\ll H\dot{\phi}$. Such limit has been studied in a cosmological context in \cite{sajf}. The Friedmann equations just reduces to one equation:
\begin{equation}
H^2={8\pi\over 3m_{pl}^2}\Lambda(t) \,, \label{e13}
\end{equation} 
with $\Lambda(t)=A+BH(t)^2$ and $A$ and $B$ constants given by
\begin{equation}
A= {m_{pl}^2\over 8\pi}\Lambda+{1\over 2}m^2\phi^2 -\frac{m^4(3-2\gamma)\hbar}{64\pi^2}\,, \hspace{0.7cm} B= 3\phi^2({1\over 8}-\xi)-3(1-2\xi)\frac{m^2(1-\gamma)\hbar}{16\pi^2}\,.\label{e14}
\end{equation}
Such behaviour is analogous to models having a time varying cosmological term, which are motivated by renormalization group to the quantum vacuum energy \cite{ren1,ren3}.  Also, analytical  solutions for the Friedmann equation (\ref{e13}) have also been obtained for phenomenological models with time varying cosmic terms \cite{sandro}. The solution for the scale factor is
\begin{equation}
a(t)=a_0 \exp\bigg({2\over 3}\sqrt{\frac{3\pi A}{3m_{pl}^2-8\pi B}}\, t\bigg)\,,
\end{equation}
indicating a de Sitter solution for the scale factor, which can indicate an accelerating solution for late time evolution. By supposing $\phi$ and $m$ much smaller than $m_{pl}$ and due to negative contribution of the quantum correction into $A$, the net effect in the evolution is to smooth the growth of the scale factor.

\section{Finite-temperature corrections and its late time cosmological consequences}\label{temperature}

The possible extension of the effective lagrangian to encompass finite-temperature effects can be obtained from the formalism just applied previously. In order to do so, we shall impose a periodicity condition on the imaginary time $y^0$ in the configuration space Green function. Then, performing the shift $\tau\rightarrow\tau+n\beta$, where $\beta=1/k_{B}T$ and $k_{B}$ is the Boltzmann constant, and summing over $n$ (see \cite{sty} and references therein) one is able to express the propagator as
\begin{eqnarray}
G_{\beta}(y,y^{\prime})= \sum_{n=-\infty}^{\infty}G(x+n\beta u,x^{\prime}),\quad u=(1,0,0,0).
\end{eqnarray}
Taking advantage of the delta distribution and the Poisson summation formula \cite{sty}
\begin{eqnarray}
\sum_{n--\infty}^{\infty}e^{ip_{0}n\beta} = \frac{2\pi}{\beta}\sum_{n=-\infty}^{\infty}\delta\left(\!p_0-\frac{2\pi}{\beta}\!\right),
\end{eqnarray}  
one obtain as the thermal Green function for the case at hand the expression
\begin{eqnarray}\label{greenthermal}
G_{\beta}^{0}(y,y^{\prime})=\frac{-i}{8\pi^{3/2}\beta}\int_{0}^{\infty}ds\frac{e^{-\alpha^2_{E}s}}{s^{3/2}}\sum_{n=-\infty}^{\infty}e^{p_{0}^2+3iH(it)p_{0}}\mathbbm{1}.
\end{eqnarray}
Inserting Eq.\eqref{greenthermal} into Eq.\eqref{loop} we obtain
\begin{eqnarray}\label{lthermal}
\mathcal{L}_{\beta}^{(1)}=-\frac{\hbar}{4\pi^{3/2}\beta}\Gamma(-3/2)\sum_{n=-\infty}^{\infty}\bigg[\alpha^{2}+\Big(\frac{2\pi n}{\beta}\Big)^2+3H(t)\frac{2\pi n}{\beta}\bigg]^{3/2}.\label{40}
\end{eqnarray}

Aiming to extract some physical information about the correction just presented, we consider the expression (\ref{lthermal}) at present time. In such a situation the term containing the Hubble parameter $H(t)/\beta$ become insignificant when compared to the previous terms. Therefore we are left with 
\begin{eqnarray}
\mathcal{L}_{\beta(0)}^{(1)}=-\frac{\hbar}{4\pi^{3/2}\beta}\Gamma(-3/2)\sum_{n=-\infty}^{\infty}\bigg[\alpha^{2}+\Big(\frac{2\pi n}{\beta}\Big)^2\bigg]^{3/2},
\end{eqnarray} where the subscript $(0)$ denotes the present time context. Looking for a finite quantum correction, we proceed manipulating the sum using methods of the regularized zeta function and dimensional regularization. Thus, after some manipulations we are able to write $\mathcal{L}^{(1)}_{\beta(0)}$ in terms of zeta function as
\begin{eqnarray}
\mathcal{L}^{(1)}_{\beta(0)}=\frac{\hbar}{3\pi^2}\bigg[\frac{32\pi^4}{\alpha\beta^5}\zeta(-3)+\frac{12\pi^2\alpha}{\beta^{3}}\zeta(-1)+\frac{\alpha^3}{\beta}\left(\frac{3}{4}\zeta(0)+1\!\right)-\frac{3\alpha^5\beta}{32\pi}\zeta(-3)+\mathcal{O}(\alpha^7\beta^3)\bigg],\label{42}
\end{eqnarray}
or, in a more direct fashion
\begin{eqnarray}
\mathcal{L}^{(1)}_{\beta(0)}=\frac{\hbar}{3\pi^2}\bigg[\frac{4\pi^4}{15\alpha\beta^5}-\frac{\pi^2\alpha}{\beta^3}+\frac{5\alpha^3}{8\beta}-\frac{\alpha^5\beta}{1280\pi}+\mathcal{O}(\alpha^7\beta^3)\bigg],\label{43}
\end{eqnarray} yielding to a finite result.

In order to maintain the approximation in which the last term of \eqref{40} is negligible today, we must impose $\alpha \sim m \gg H_0 \sim 10^{-33}$eV. Also, for recent times, where $\beta_0\equiv 1/k_BT_0\sim 10^4$eV$^{-1}$, with $T_0\sim 2.7$K, the first term of (\ref{43}) dominates if $\alpha \sim m \ll 1/\beta_0 \sim 10^{-4}$eV, which puts an upper limit to the mass of the field in order to have a finite sum from \eqref{43}.  With such approximation, we have the mass of the field constrained to $H_0 \ll m \ll 1/\beta_0$, which corresponds to $ 10^{-33}$eV$\ll m \ll 10^{-4}$eV. 

In such limit the quantum correction we are interested is dominated by the first term of \eqref{43}, which must accompany the potential term of the bare lagrangian, namely the term $m^2\phi^2$, which shall corrects the Friedmann equation accordingly. Thus, looking for the present time slowly varying limit of \eqref{e13}, we have the term $A$ corrected to
\begin{equation}
A'= {m_{pl}^2\over 8\pi}\Lambda+{1\over 2}m^2\phi^2 -\frac{m^4(3-2\gamma)\hbar}{64\pi^2} + \frac{ 4\pi^2 \hbar}{45 m \beta_0^5}\,. \label{44}
\end{equation}
With the upper limit to the mass imposed above and supposing $\phi \sim m$, the last term of \eqref{44} dominates over the second and third one, and additionally $A'\gg BH_0^2$ from \eqref{e14}, showing that the quantum correction at finite temperature act exactly like a cosmological constant term, evolving as
\begin{equation}
a(t)=a_0 \exp\bigg(\sqrt{\frac{32\pi^2 \hbar}{135m_{pl}^2m\beta_0^5}}\, t\bigg)\,,
\end{equation}
which could drive the recent phase of acceleration of the universe. The zero temperature limit corresponds to $\beta_0 \to \infty$, which cancels the contribution to accelerated expansion due to temperature effects. Also, for larger values of the mass $m$ of the field (in the range above) the expansion is attenuated, showing the effect of the gravitational attraction against the repulsion. Finally, in order to reproduce the expected value of $10^{-47}$GeV$^{-4}$ for the energy density of the cosmological constant according to standard $\Lambda$CDM model, the mass of the field must be $m\simeq 10^{-9}$eV, in good agreement to the limit range adopted above.

\section{Final Remarks}\label{finalrem}
In this work we completed the program of deriving the effective lagrangian for a mass-dimension-one fermionic field in a curved space-time with slowly varying background field in a quasi-local situation. We also have computed the one-loop corrections in the finite temperature situation. 

More than an academic exercise, we highlight some cosmological applications of the present study, {at least in the cases of smooth and adiabatic background evolution}. In the zero temperature case, we have analyzed two different cases, corresponding to an early time universe and a late time universe. In the zero temperature limit we have seen that in the limit $H\approx m_{pl} \gg m \gg \phi$ and $\xi>1/2$ a positive quantum cosmological term appears naturally, which may be responsible for the accelerated inflationary expansion after about $t\sim H^{-1}\sim 10^{-43}$s, where the quantum effects are to be relevant. For the late time evolution we studied the limit $\dot{\phi}\ll H\phi$, $\ddot{\phi}\ll H\dot{\phi}$ and $\phi \approx constant$, which corresponds to a model with a time varying cosmological term already studied in \cite{sajf}, but here with the corresponding quantum correction, whose solution for the scale factor is also of accelerating type, an exponential growth. A most complete model should, eventually, also include additional matter components, as radiation and baryonic matter.

In the case of finite temperature corrections, the cosmological scenery studied was that one corresponding to late time expansion with low temperature limit and also $m\sim \phi \gg H_0$,  leading to an interesting contribution of the temperature correction acting like a cosmological constant term, and setting a limit to the mass of the field of about $10^{-9}$eV in order to reproduce the value of the the standard model. This contribution comes exclusively from the finite temperature correction. 

\section{ACKNOWLEDGMENT}
RJBR would like to thanks Dr. Carlos Hugo Coronado Villalobos for useful discussions during the manuscript writing stage and to CAPES for the financial support. SHP is grateful to CNPq - Conselho Nacional de Desenvolvimento Cient\'ifico e Tecnol\'ogico, Brazilian research agency, for financial support (No. 304297/2015-1; 400924/2016-1). JMHS thanks to CNPq for financial support (No. 304629/2015-4; 445385/2014-6).

\end{document}